\documentclass[twocolumn]{aastex7}
\usepackage{amsmath,amstext}
\usepackage[T1]{fontenc}
\usepackage{apjfonts} 
\usepackage{natbib}
\usepackage{longtable}
\usepackage{comment}

\newcommand{\jwst}{\textit{JWST}}

\newcommand{\ha}{\hbox{{\rm H}$\alpha$}}
\newcommand{\Hb}{\hbox{{\rm H}$\beta$}}
\newcommand{\hb}{\hbox{{\rm H}$\beta$}}

\newcommand{\heii}{\hbox{\ion{He}{2}}}

\newcommand{\nii}{\hbox{[\ion{N}{2}]}}

\newcommand{\oi}{\hbox{[\ion{O}{1}]}}
\newcommand{\oii}{\hbox{[\ion{O}{2}]}}

\newcommand{\oiii}{\hbox{[\ion{O}{3}]}}

\newcommand{\neiii}{\hbox{[\ion{Ne}{3}]}}

\newcommand{\nev}{\hbox{[\ion{Ne}{5}]}}

\newcommand{\sii}{\hbox{[\ion{S}{2}]}}

\newcommand{\mbh}{\hbox{$M_\mathrm{BH}$}}
\newcommand{\msun}{\hbox{$M_\odot$}}

\newcommand{\zsol}{\hbox{$Z_\odot$}}

\newcommand{\zgas}{\hbox{Z$_\mathrm{gas}$}}

\newcommand{\qtotal}{\hbox{$Q_\mathrm{total}$}}
\newcommand{\qlow}{\hbox{$Q_\mathrm{low}$}}
\newcommand{\qint}{\hbox{$Q_\mathrm{intermediate}$}}
\newcommand{\qhigh}{\hbox{$Q_\mathrm{high}$}}
\newcommand{\qveryhigh}{\hbox{$Q_\mathrm{very~high}$}}

\begin{document}

\title{\large \bf Optical Strong Line Ratios Cannot Distinguish Between Stellar Populations and Accreting Black Holes at High Ionization Parameters and Low Metallicities}

\correspondingauthor{Nikko J. Cleri}
\email{cleri@psu.edu}


\author[0000-0001-7151-009X]{Nikko J. Cleri}
\affiliation{Department of Astronomy and Astrophysics, The Pennsylvania State University, University Park, PA 16802, USA}
\affiliation{Institute for Computational and Data Sciences, The Pennsylvania State University, University Park, PA 16802, USA}
\affiliation{Institute for Gravitation and the Cosmos, The Pennsylvania State University, University Park, PA 16802, USA}
\affiliation{Department of Physics and Astronomy, Texas A\&M University, College Station, TX, 77843-4242 USA}
\affiliation{George P.\ and Cynthia Woods Mitchell Institute for Fundamental Physics and Astronomy, Texas A\&M University, College Station, TX, 77843-4242 USA}
\email{cleri@psu.edu}

\author[0000-0002-4606-4240]{Grace M. Olivier}
\affiliation{Department of Physics and Astronomy, Texas A\&M University, College Station, TX, 77843-4242 USA}
\affiliation{George P.\ and Cynthia Woods Mitchell Institute for Fundamental Physics and Astronomy, Texas A\&M University, College Station, TX, 77843-4242 USA}
\email{gmolivier@tamu.edu}

\author[0000-0001-8534-7502]{Bren E. Backhaus}
\affil{Department of Physics and Astronomy, University of Kansas, Lawrence, KS 66045, USA}
\email{bren.backhaus@ku.edu}

\author[0000-0001-6755-1315]{Joel Leja}
\affiliation{Department of Astronomy and Astrophysics, The Pennsylvania State University, University Park, PA 16802, USA}
\affiliation{Institute for Computational and Data Sciences, The Pennsylvania State University, University Park, PA 16802, USA}
\affiliation{Institute for Gravitation and the Cosmos, The Pennsylvania State University, University Park, PA 16802, USA}
\email{joel.leja@psu.edu}

\author[0000-0001-7503-8482]{Casey Papovich}
\affiliation{Department of Physics and Astronomy, Texas A\&M University, College
Station, TX, 77843-4242 USA}
\affiliation{George P.\ and Cynthia Woods Mitchell Institute for
 Fundamental Physics and Astronomy, Texas A\&M University, College
 Station, TX, 77843-4242 USA}
\email{papovich@tamu.edu}

\author[0000-0002-1410-0470]{Jonathan R. Trump}
\affil{Department of Physics, 196A Auditorium Road, Unit 3046, University of Connecticut, Storrs, CT 06269, USA}
\email{jonathan.trump@uconn.edu}

\author[0000-0002-7959-8783]{Pablo Arrabal Haro}
\altaffiliation{NASA Postdoctoral Fellow}
\affiliation{Astrophysics Science Division, NASA Goddard Space Flight Center, 8800 Greenbelt Rd, Greenbelt, MD 20771, USA}
\email{pablo.arrabalharo@nasa.gov}

\author[0000-0003-3441-903X]{V\'eronique Buat}
\affiliation{Aix Marseille Univ, CNRS, CNES, LAM Marseille, France}
\email{veronique.buat@lam.fr}

\author[0000-0002-4193-2539]{Denis Burgarella}
\affiliation{Aix Marseille Univ, CNRS, CNES, LAM Marseille, France}
\email{denis.burgarella@lam.fr}

\author[0000-0001-8174-317X]{Emilie Burnham}
\affiliation{Department of Astronomy and Astrophysics, The Pennsylvania State University, University Park, PA 16802, USA}
\affiliation{Institute for Gravitation and the Cosmos, The Pennsylvania State University, University Park, PA 16802, USA}
\email{efb555@@psu.edu}

\author[0000-0003-2536-1614]{Antonello Calabr\`o}
\affiliation{INAF Osservatorio Astronomico di Roma, Via Frascati 33, 00078 Monte Porzio Catone, Rome, Italy}
 \email{antonello.calabro@inaf.it}

 \author[0000-0003-1420-6037]{Jonathan H. Cohn}
 \affiliation{Department of Physics and Astronomy, Dartmouth College, 6127 Wilder Laboratory, Hanover, NH 03755, USA}
 \email{Jonathan.Cohn@dartmouth.edu}

\author[0000-0002-6348-1900]{Justin W. Cole}
\altaffiliation{NASA FINESST Fellow}
\affiliation{Department of Physics and Astronomy, Texas A\&M
  University, College Station, TX, 77843-4242 USA}
\affiliation{George P.\ and Cynthia Woods Mitchell Institute for
  Fundamental Physics and Astronomy, Texas A\&M University, College
  Station, TX, 77843-4242 USA}
\email{jwc68@tamu.edu}

\author[0000-0001-8047-8351]{Kelcey Davis}
\altaffiliation{NSF Graduate Research Fellow}
\affiliation{Department of Physics, 196A Auditorium Road, Unit 3046, University of Connecticut, Storrs, CT 06269, USA}
\email{kelcey.davis@uconn.edu}

\author[0000-0001-5414-5131]{Mark Dickinson}
\affiliation{NSF's National Optical-Infrared Astronomy Research Laboratory, 950 N. Cherry Ave., Tucson, AZ 85719, USA}
\email{mark.dickinson@noirlab.edu}

 \author[0000-0001-8519-1130]{Steven L. Finkelstein}
\affiliation{Department of Astronomy, The University of Texas at Austin, Austin, TX, USA}
\email{stevenf@astro.as.utexas.edu}

\author[0000-0002-9426-7456]{Ray Garner, III}
\affiliation{Department of Physics and Astronomy, Texas A\&M University, College Station, TX, 77843-4242 USA}
\affiliation{George P.\ and Cynthia Woods Mitchell Institute for Fundamental Physics and Astronomy, Texas A\&M University, College Station, TX, 77843-4242 USA}
\email{ray.three.garner@gmail.com}

\author[0000-0002-3301-3321]{Michaela Hirschmann}
\affiliation{Institute of Physics, Laboratory of Galaxy Evolution, Ecole Polytechnique Federale de Lausanne (EPFL), Observatoire de Sauverny, 1290 Versoix, Switzerland}
\email{michaela.hirschmann@epfl.ch}

\author[0000-0003-3424-3230]{Weida Hu}
\affiliation{Department of Physics and Astronomy, Texas A\&M
  University, College Station, TX, 77843-4242 USA}
\affiliation{George P.\ and Cynthia Woods Mitchell Institute for
  Fundamental Physics and Astronomy, Texas A\&M University, College
  Station, TX, 77843-4242 USA}
\email{weidahu@tamu.edu}

\author[0000-0001-6251-4988]{Taylor A. Hutchison}
\altaffiliation{NASA Postdoctoral Fellow}
\affiliation{Astrophysics Science Division, NASA Goddard Space Flight Center, 8800 Greenbelt Rd, Greenbelt, MD 20771, USA}
\email{taylor.hutchison@nasa.gov}

\author[0000-0002-8360-3880]{Dale D. Kocevski}
\affiliation{Department of Physics and Astronomy, Colby College, Waterville, ME 04901, USA}
\email{dkocevski@colby.edu}

\author[0000-0002-6610-2048]{Anton M. Koekemoer}
\affiliation{Space Telescope Science Institute, 3700 San Martin Drive, Baltimore, MD 21218, USA}
\email{koekemoer@stsci.edu}

\author[0000-0003-2366-8858]{Rebecca L. Larson}
\affiliation{Space Telescope Science Institute, 3700 San Martin Drive, Baltimore, MD 21218, USA}
\email{rlarson@stsci.edu}

\author[0009-0009-4899-6242]{Zach J. Lewis}
\affiliation{Department of Astronomy, University of Wisconsin-Madison, Madison, WI 53706, USA}
\email{zjlewis@wisc.edu}

\author[0000-0003-0695-4414]{Michael V. Maseda}
\affiliation{Department of Astronomy, University of Wisconsin-Madison, Madison, WI 53706, USA}
\email{maseda@astro.wisc.edu}

\author[ 0000-0001-7755-4755]{Lise-Marie Seill\'e}
\affiliation{ 
Aix Marseille Univ, CNRS, CNES, LAM, Marseille, France
}
\email{lise-marie.seille@lam.fr}

\author[0000-0002-6386-7299]{Raymond C.\ Simons}
\affiliation{Department of Engineering and Physics, Providence College, 1 Cunningham Sq, Providence, RI 02918 USA}
\email{rcsimons@providence.edu}

\begin{abstract}
High-redshift observations from JWST indicate that optical strong line ratios do not carry the same constraining power as they do at low redshifts. Critically, this prevents a separation between stellar- and black hole-driven ionizing radiation, thereby obscuring both active galactic nuclei demographics and star formation rates. To investigate this, we compute a large suite of photoionization models from Cloudy powered by stellar populations and accreting black holes over a large grid of ages, metallicities, initial mass functions, binarity, ionization parameters, densities, and black hole masses. We use these models to test three rest-frame optical strong line ratio diagnostics which have been designed to separate ionizing sources at low redshifts ($z\lesssim2$): the \nii-BPT, VO87, and OHNO diagrams. We show that the position of a model in these diagrams is strongly driven by the ionization parameter (log U) and the gas-phase metallicity (\zgas), often more so than the ionizing spectrum itself; in particular, there is significant overlap between stellar population and accreting black hole models at high log U and low \zgas. We show that the OHNO diagram is especially susceptible to large contamination of the AGN region defined at $z\sim1$ for stellar models with high log U and low \zgas, consistent with many observed JWST spectra at high redshift. We show that the optical line ratio diagnostics are most sensitive to the shape of the $<$54 eV ionizing continuum, and that the derived ionizing sources for a given set of optical strong line ratios can be highly degenerate. Finally, we demonstrate that very high ionization (>54 eV) emission lines that trace ionizing sources harder than normal stellar populations help to break the degeneracies present when using the strong line diagnostics alone, even in gas conditions consistent with those at high redshifts. 
\end{abstract}

\section{Introduction} \label{sec:intro}
Reliably identifying the sources of ionizing photons is critical to properly interpreting observations of galaxies, especially at high redshifts now accessible with \jwst. Misattributed emission from accreting black holes can be responsible for orders of magnitude discrepancies of derived quantities such as star formation rates and stellar masses \citep[e.g.,][]{Brinchmann2004,Netzer2007,Rosario2013,Davies2014a,Davies2014c,Rosario2015,Shimizu2015,Magliocchetti2016,Ellison2016,Cairos2022,Wang2024,Wang2025,Siudek2025,Berger2025}. 

There exist many different methods for constraining the presence a luminous accreting black hole; objects for which the data support the presence of a luminous accreting black hole are referred to as ``active'' (or ``active galactic nuclei''; AGN), and those for which the data do not have sufficient evidence are referred to as ``inactive'' or ``star forming''. These methods are thought to be highly dependent on viewing angle \citep[e.g.,][]{Urry1995}, and range across the electromagnetic spectrum from detections of high-energy photons in the X-ray regime \citep[e.g.,][]{Xue2011,Xue2016,Luo2017}, to synchrotron emission in the radio \citep[e.g.,][]{Heckman2014,Padovani2017}. In the infrared, photometric selections have long been used to distinguish between accreting supermassive black holes and star-forming galaxies by tracing emission from hot dust largely associated with AGN \citep[e.g.,][]{Lacy2004,Donley2012,Stern2005,Kirkpatrick2017,Kirkpatrick2023}. 

Rest-frame optical diagnostics of ionizing sources are among the most widely used, largely due to the development of instruments designed to target rest-frame optical spectroscopy at various redshifts \citep[e.g.,][]{York2000,McLean2012,Jakobsen2022}. Emission lines from permitted transitions, most commonly the Balmer lines, with broad profiles (FWHM $\gtrsim 1000~\mathrm{km~s}^{-1}$) trace the dense, rapidly moving gas which is predominantly associated with the broad line region around a black hole, and are common diagnostic tools given the lack of other known physical conditions to produce such signatures \citep[e.g.][]{Osterbrock1986,Sulentic2000,Kauffmann2003,Trump2011}. 

The diagnostics that we focus on for the remainder of this work are those based on intensity ratios of rest-frame optical emission lines. Emission line ratios have been used to classify ionizing sources for several decades; early studies used only a single emission line ratio (e.g., $\oiii~\lambda5008$/\hb) to characterize a source as a star forming region or host of an accreting black hole \citep[e.g.,][]{Searle1971,Smith1975,Alloin1978,Shuder1981}. However, it was quickly shown that this one-dimensional determination was insufficient for separating Seyfert 2s from intense starburst galaxies or shock heating \citep[e.g.,][]{Heckman1980,Baldwin1981,Balzano1983,Keel1983,Osterbrock1985}.

The addition of a second dimension with another emission line ratio led to the ``BPT'' diagrams\footnote{Colloquially, the terms ``BPT'' or ``BPT-style/BPT-like'' have been used to refer to any emission line ratio diagnostics used to separate ionizing sources. For clarity, we will refer to each of the three diagrams studied in this work explicitly; the \oiii/\hb\ versus \nii/\ha\ diagram as the ``\nii-BPT'' diagram, the \oiii/\hb\ versus \sii/\ha\ diagram as the ``VO87'' diagram, and the \oiii/\hb\ versus \neiii/\oii\ diagram as the ``OHNO'' diagram.}, which compared $\oiii~\lambda5008$/\hb\ to $\nii~\lambda6585$/\ha\ or $\oi~\lambda6302$/\ha\ \citep{Baldwin1981}. These diagnostics were designed with the criteria: (1) the line ratios were made up of strong, easily detectable lines, (2) lines that are significantly blended should be avoided, (3) wavelength separation should be small to limit effects of dust attenuation and instrumental calibrations, (4) ratios of a forbidden line to a Balmer line are preferred to limit abundance sensitivity, and (5) lines used in these diagnostics should all be readily accessible with current instrumentation \citep{Veilleux1987}. 

The \nii-BPT diagnostic utilizes the ratios \oiii/\hb\ vs. \nii/\ha\ \citep{Baldwin1981,Kauffmann2003,Kewley2001,Kewley2006,Trump2015}. The original samples used to test the \nii-BPT diagnostic in \cite{Baldwin1981} included local-Universe planetary nebulae, Seyfert 1s and 2s, Narrow-Line Radio Galaxies, Herbig-Haro objects, shock heated galaxies, and galactic and detached extragalactic H II regions. The original \cite{Baldwin1981} analysis does not make a strict demarcation for the locations of different physical sources, though it is noted that H II regions are generally lower in \oiii/\hb\ and \nii/\ha\ than sources ionized by other physical mechanisms (e.g., planetary nebulae and accreting black holes). 

Using the photoionization modeling code MAPPINGS III \citep{Sutherland1993} to model the extent of infrared starburst galaxies in emission line ratio diagnostic space, \cite{Kewley2001} defines the extreme starburst classification on the \nii-BPT diagram as
\begin{align}\label{eq:bpt_kewley}
    \log\left(\frac{\oiii}{\Hb}\right)&=\frac{0.61}{\log\left(\frac{\nii}{\ha}\right)-0.47} + 1.19
\end{align}
where objects with line ratios above this curve are denoted as AGN, and below this curve are denoted as star-forming. 

Studies of large statistical samples of galaxies from the Sloan Digital Sky Survey \citep{York2000} tested the \nii-BPT diagram and derived now-widely used diagnostics to separate H II regions from regions powered by accreting black holes. \cite{Kauffmann2003} defines the pure star formation classification
\begin{align}\label{eq:bpt_kauffmann}
    \log\left(\frac{\oiii}{\Hb}\right)&=\frac{0.61}{\log\left(\frac{\nii}{\ha}\right)-0.05} + 1.3
\end{align}
where objects with line ratios above this curve are denoted as AGN, and below this curve are denoted as star-forming. 

These two classifications are designed to probe different physical regimes of ionizing sources: the maximal extent of star formation (for the \cite{Kewley2001} diagnostic) and the region of highest purity for star formation (for the \cite{Kauffmann2003} diagnostic). As such, the location of a source on the \oiii/\hb\ vs. \nii/\ha\ plane relative to these demarcations should be interpreted with their specific goals and biases in mind. Combining these two diagnostics yields a third region, that in between the two divisions, which has become known as the ``composite'' region \citep{Kewley2006}. 

Similarly to the \nii-BPT, \cite{Veilleux1987} introduced the $\oiii ~\lambda5008$/\hb\ vs. $\sii ~\lambda\lambda6718,6733$/\ha\ diagram (dubbed ``VO87'') using the criteria (1) each line ratio should be made up of strong lines, (2) lines that suffer severe blending should be avoided, (3) lines should be close in wavelength to mitigate the effects of attenuation and flux calibrations, (4) ratios of a line of one element to a Balmer line are preferred to limit abundance sensitivity, and (5) lines should be accessible to prevalent instruments (at the time of \cite{Veilleux1987}, this indicated a preference for optical over UV or other wavelength lines). The choice of \sii/\ha\ is made to alleviate potential blending of \nii\ and \ha\ in the traditional \nii-BPT diagram, which can be an issue at lower spectral resolutions. \cite{Veilleux1987} does not give the division between H II region galaxies and AGNs quantitatively for the VO87 diagram, but they do note that there are decisive regions populated by HII galaxies (low \oiii/\hb\ and low \sii/\ha) and AGN (high \oiii/\hb\ and high \sii/\ha). 

Other works use models and large statistical samples to quantify the VO87 diagnostic, including \cite{Kewley2001,Kewley2006} with the maximal starburst line
\begin{align}\label{eq:vo_kewley}
    \log\left(\frac{\oiii}{\Hb}\right)&=\frac{0.72}{\log\left(\frac{\sii}{\ha}\right)-0.32} + 1.3
\end{align}
and the division between AGN and low-ionization nuclear emission-line regions (LINERs)
\begin{align}\label{eq:vo_kewley_liner}
    \log\left(\frac{\oiii}{\Hb}\right)&=\frac{01.89}{\log\left(\frac{\sii}{\ha}\right)} + 0.76
\end{align}
where galaxies in the low \oiii/\hb\ and low \sii/\ha\ sector are classified as star-forming, those in the low \oiii/\hb\ and high \sii/\ha\ sector are classified as LINERs/shocks, and those in the high \oiii/\hb\ region are classified as AGN.

Another diagnostic for the VO87 diagram was introduced in \cite{Trump2015}, which is fashioned similarly to the \cite{Kauffmann2003} \nii-BPT line, and differs from the \cite{Kewley2001,Kewley2006} maximal starburst line in the treatment of low metallicity H II regions, which tend to live in the high \oiii/\hb\ and low \sii/\ha\ region of VO87. The \cite{Trump2015} diagnostic is defined as
\begin{align}\label{eq:vo_trump}
    \log\left(\frac{\oiii}{\Hb}\right)&=\frac{0.48}{\log\left(\frac{\sii}{\ha}\right)-0.10} + 1.3
\end{align}

Several line ratio diagnostics have modified the criteria of the \nii-BPT and VO87 diagrams in efforts to produce greater purity and/or completeness. Some versions leverage line ratios of nearby forbidden lines \citep[e.g.,][]{Cleri2023b,Mazzolari2024,Backhaus2025}, line ratios at wavelengths other than the rest-frame optical \citep[e.g.][]{Allen1998,Nakajima2018,Hirschmann2019,Hirschmann2023,Mingozzi2024,Flury2024,Calabro2023,Petric2011,Feltre2023}, or forego a line ratio in favor of a derived quantity, such as stellar mass \citep[e.g., the ``Mass-Excitation diagram'',][]{Juneau2011,Juneau2014,Coil2015}, rest frame optical and near-IR colors \citep[e.g.,][]{Yan2011,Trouille2011}, spectral break strengths \citep[e.g.,][]{Stasinska2006}, equivalent widths of individual lines \citep[e.g.,][]{CidFernandes2010}, or include spatially-resolved kinematics from integral field spectroscopy \citep[e.g.,][]{Zhu2025}. 

Developed recently compared to the \nii-BPT and VO87 diagrams, the \oiii/\hb\ vs. $\neiii ~\lambda3870$/$\oii~\lambda\lambda3727,3730$ diagram (dubbed ``OHNO'') was designed to separate Chandra X-ray selected AGN from other sources at $z\sim1$ \citep{Zeimann2015,Backhaus2022,Cleri2023a}. \neiii/\oii\ has been used as an ionization indicator on its own \citep[e.g.,][]{Levesque2014}: the ionization energy needed to produce \neiii\ is 40.96 eV (slightly higher than that of \oiii\ at 35.12 eV), and the ionization energy needed to produce \oii\ is 13.62 eV, so the ratio traces ionization slightly harder than \oiii/\hb. 

The OHNO diagram utilizes \neiii\ and \oii, which are bluer than the \ha, \nii, and \sii\ complex, thus giving OHNO the potential to be used at higher redshifts. Additionally, the use of oxygen and neon lines removes the N/O abundance sensitivity of the \nii-BPT arising from the primary and secondary production pathways of nitrogen \citep[e.g.,][]{Henry2000}. OHNO also comes with several downsides compared to the \nii-BPT and VO87. First, \neiii/\oii\ fails the criterion of the \nii-BPT and VO87 diagrams which prefer the ratios of forbidden metal lines to Balmer lines. More critically, the fundamental physical mechanism probed by both axes of OHNO is the ionization parameter, and the similarities of the ionization energies probed by each ratio does not offer a truly orthogonal dimension in which to separate ionizing sources \citep[e.g.,][]{Backhaus2024,Backhaus2025,Cleri2023a,Cleri2023b,Larson2023}.

The OHNO diagnostic is defined by \cite{Backhaus2022} as 
\begin{align}\label{eq:ohno}
\log\left(\frac{\oiii}{\Hb}\right)=\frac{0.35}{2.8\log( \neiii/\oii) - 0.8} + 0.64
\end{align}
where it was used to show that X-ray AGN \citep[as defined by][]{Xue2011,Xue2016,Luo2017} and galaxies with very-high ionization emission lines (particularly $\nev~\lambda 3427$) in their spectra are preferentially classified as AGN near the peak of cosmic star formation and AGN activity at $z\sim 1-2$ \citep{Backhaus2022,Cleri2023a}. 

A common issue with most of these optical line ratio diagnostics is the lack of systematic testing through observations at high redshifts ($z > 2$).  It has been shown extensively that there is evolution in the gas conditions and stellar populations in galaxies out to $z\sim2$ \citep[e.g.,][see Section \ref{sec:discussion} for a full discussion]{Sanders2015,Sanders2016,Shapley2015,Strom2017,Papovich2022}, where the lower metallicities, higher densities, and harder ionizing radiation from star forming galaxies may have a significant impact on the classification of ionizing sources \citep[e.g.,][]{Baldwin1981,Veilleux1987,Kewley2006,Juneau2011,Juneau2014,Kewley2013,Coil2015}. The lack of study beyond $z\sim2$ is primarily due to the lack of instrumentation capable of producing high-quality rest-frame optical spectra in these early regimes prior to JWST.

With the spectroscopic capabilities and longer wavelength coverage of instruments like JWST/NIRSpec \citep{Jakobsen2022}, we are now able to probe the ionizing properties of early galaxies with sample sizes unmatched prior to JWST. The optical line ratio diagnostics are convenient and accessible up to $z\sim9$, where \oiii\ is redshifted out of of the spectral coverage of NIRSpec. 

Given the ease of use for high redshift spectra, several recent works have used the \nii-BPT, VO87, and OHNO diagrams at $z>>2$ with JWST spectroscopy. These studies have shown the potential for large amounts of contamination in both the star forming and AGN regions of these diagrams due to the evolving behavior of stellar populations and gas conditions across cosmic time \citep[e.g.,][see Section \ref{sec:discussion} for a full discussion]{Cleri2023b,Ubler2023,Larson2023,Scholtz2025,Sanders2023,Cameron2023a}. This motivates a systematic analysis of optical line ratio diagnostics through observations in forthcoming work (Cleri et al. in preparation), and a reanalysis of the existing diagnostics of ionizing sources and their efficacy at high redshifts in this work.

In this work, we compute a library of photoionization models to test the utility of these optical strong line ratio diagnostics in preparation for large statistical samples of JWST observations at high redshifts. The remainder of this paper is structured as follows. In Section \ref{sec:models}, we discuss the framework of our photoionization modeling. In Section \ref{sec:results}, we analyze the \nii-BPT, VO87, and OHNO diagrams in the context of our models. In Section \ref{sec:discussion}, we discuss the implications of our results on the future use of emission line ratio diagnostics, particularly in the context of spectroscopy of sources in the early Universe. In Section \ref{sec:conclusions}, we summarize the results and conclusions of this analysis. 

\section{Photoionization Models}\label{sec:models}
For the following analysis, we use Cloudy version C23.01 \citep{Gunasekera2023,Ferland2017}. Cloudy is a photoionization simulation code designed to self-consistently model physical conditions in astrophysical clouds to predict thermal, ionization, and chemical structure of the cloud and predict its observed spectrum. Along with this paper we publicly release the extended Cloudy model library\footnote{The full Cloudy model library is currently available for download at \href{https://njcleri.github.io/products.html}{https://njcleri.github.io/products.html} or by request to the corresponding author.} which includes emission line emissivities and  continua for all stellar parameters and includes additional runs for varying abundance patterns which are not included in this work.

We perform our photoionization modeling using Cloudy on the Texas A\&M High Performance Research Computing (HPRC) Grace cluster\footnote{See \url{https://hprc.tamu.edu/kb/User-Guides/Grace}, where the namesake of the cluster is  \href{https://en.wikipedia.org/wiki/Grace_Hopper}{Grace Hopper}.}, which has 800 Intel 6248R 3.0GHz 24-core processors with (at least) 384 Gb RAM. On a single core, the time to run a single model is of order $\sim$1 minute. 

We compute all of our models across a grid of ionization parameters\footnote{This is the initial ionization parameter Cloudy assumes at the incident face of the cloud.} from $-4<\log U<-1$ in steps of 0.25, where \cite{Kewley2019b} defines the dimensionless ionization parameter U as 
\begin{align}\label{eq:U}
    U &\equiv \frac{1}{c}\frac{\Phi}{n_H}
\end{align}
where $\Phi$ is the ionizing photon flux and $n_H$ is the hydrogen density. We assume a plane-parallel gas geometry for all models, which is an unphysical but necessary simplification over a generalized geometry for a model grid of this size \citep[][]{Katz2023c}. 

The details of the individual models used in this work are described in the following subsections.

\subsection{BPASS Stellar Population Models}
The stellar population models used in this work are from the Binary Population and Spectral Synthesis (BPASS; v2.2.1) single-burst binary-formation library \citep{Stanway2018}. BPASS v2.2.1 has nine different initial mass functions (IMFs), which include four different high-mass slopes and high-mass cutoffs of 100 or 300 \msun\ \citep[where the 135\_300 corresponds to the BPASS default IMF,][]{Salpeter1955}. In this work, we compute nebular line emissivities for the IMFs listed in Table \ref{tab:bpass_imfs} \citep[for more details, see Section 2.4 and Table 1 of][]{Stanway2018}. 

\begin{deluxetable}{l|cccccccc}[t]
\tablecaption{BPASS IMFs tested in this work. } \label{tab:bpass_imfs}
\tablehead{
\colhead{BPASS IMF} & \colhead{$\alpha_1$} & \colhead{$\alpha_2$} & \colhead{$M_1$ [\msun]} & \colhead{$M_{max}$ [\msun]}
}
\startdata
    100\_300 & -1.30 & -2.00 & 0.5 & 300 \\
    135\_300 & -1.30 & -2.35 & 0.5 & 300 \\
    170\_300 & -1.30 & -2.70 & 0.5 & 300 \\
    chab300 & \textit{exp. cutoff} & -2.30 & 1.0 & 300 \\
\enddata
\end{deluxetable}

BPASS offers thirteen different metallicities: $10^{-5}$, $10^{-4}$, $10^{-3}$, 0.002, 0.003, 0.004, 0.005, 0.006, 0.008, 0.010, 0.014, 0.020, 0.040, where it is taken that $0.020 = \zsol$ \citep[recent works have suggested that $0.0225 = \zsol$, though we maintain the BPASS convention throughout this work;][]{Magg2022}. BPASS offers a grid of ages from $\log(\mathrm{age/yr}) = 6.0$ to $\log(\mathrm{age/yr}) = 11.0$ in steps of $\log(\mathrm{age/yr}) = 0.1$. To optimize computation time, we test stellar ages in a grid of step size $\log(\mathrm{age/yr}) = 0.1$ from $\log(\mathrm{age/yr}) = 6.0$ to $\log(\mathrm{age}) = 8.0$ and $\log(\mathrm{age/yr}) = 0.5$ from $\log(\mathrm{age/yr}) = 8.0$ to $\log(\mathrm{age/yr}) = 11.0$, as stars of these ages in the BPASS models do have a significant ionizing photon continuum \citep{Stanway2018}.  We test both the single star and binary star populations. We use the single-burst star formation histories for all BPASS models. We assume the Cloudy default \cite{Grevesse2010} solar abundance ratios and Orion dust grains for the initial gas phase and dust abundances, with nebular metallicity scaled directly with stellar metallicity. We compute our stellar models at densities $n_H = 10^2,~10^3 ~\mathrm{cm}^{-3}$.







\subsection{Black Hole Accretion Models}

The models of black hole accretion used in this work stem from the SEDs of \cite{Done2012}. These SEDs include a prescription for a blackbody accretion disk modeled as a series of blackbodies of varying temperatures as well as a Compton upscattering component from the disk, which recreates the ``soft X-ray excess'' often observed in local Universe sources \citep[e.g.,][]{Done2012,Gierlinski2004,Walter1993,Mehdipour2015}. The high energy emission results from a second Compton upscattering in an optically thin corona above the disk, modeled by a power law tail to the SED. 

The physical parameters of the black hole accretion disk models are the \texttt{XSPEC} \citep{Arnaud1996} defaults except where specified: The black hole mass, \mbh, is varied over $\log\mbh/\msun \in \{3,4,5,6,7,8,9\}$. The Eddington ratio is $\log L_{bol}/L_{Edd} = -1$. The corona radius is set to 10$R_g$ where the gravitational radius is $R_g \equiv G\mbh/c^2$. The outer radius of the accretion disk is set to $\log r_{out}/R_g = 5$. The soft Comptonization component is set to $kT_e = 0.1$ keV, which is different from the \texttt{XSPEC} default but is chosen to maintain consistency with previous studies \citep{Cann2018}. The optical depth of the soft Comptonization component is set to $\tau=10$. The spectral index of the hard Comptonization component is set to 10. The fraction of the power law below $r_{cor}$ which is emitted in the hard component is set to $\log f_{pl} = -4$. The SEDs can be retrieved from \texttt{XSPEC} \citep{Arnaud1996} using the \textsc{optxagnf} command. We compute our black hole accretion disk models at densities $n_H = 10^2,~10^3,~10^4 ~\mathrm{cm}^{-3}$.

The true mechanics of the emission from black hole accretion disks are highly intricate, far more so than the relatively simplistic models presented in this work \citep[e.g.,][see Section \ref{subsec:caveats} for more discussion]{Adhikari2016,Mitchell2023,McKaig2023,Cann2018}. Our models assume that the accretion disk ionizes a single cloud. This simplification is reasonable for this work, as models with separate BLR and NLR physics are most likely to impact higher energy regimes than the emission lines focused on here (i.e., $\lesssim$100eV) \citep[e.g.,][]{McKaig2023}. We discuss the caveats and limitations of these models in Section \ref{subsec:caveats}.


\begin{figure*}[t]
\epsscale{1.1}
\plotone{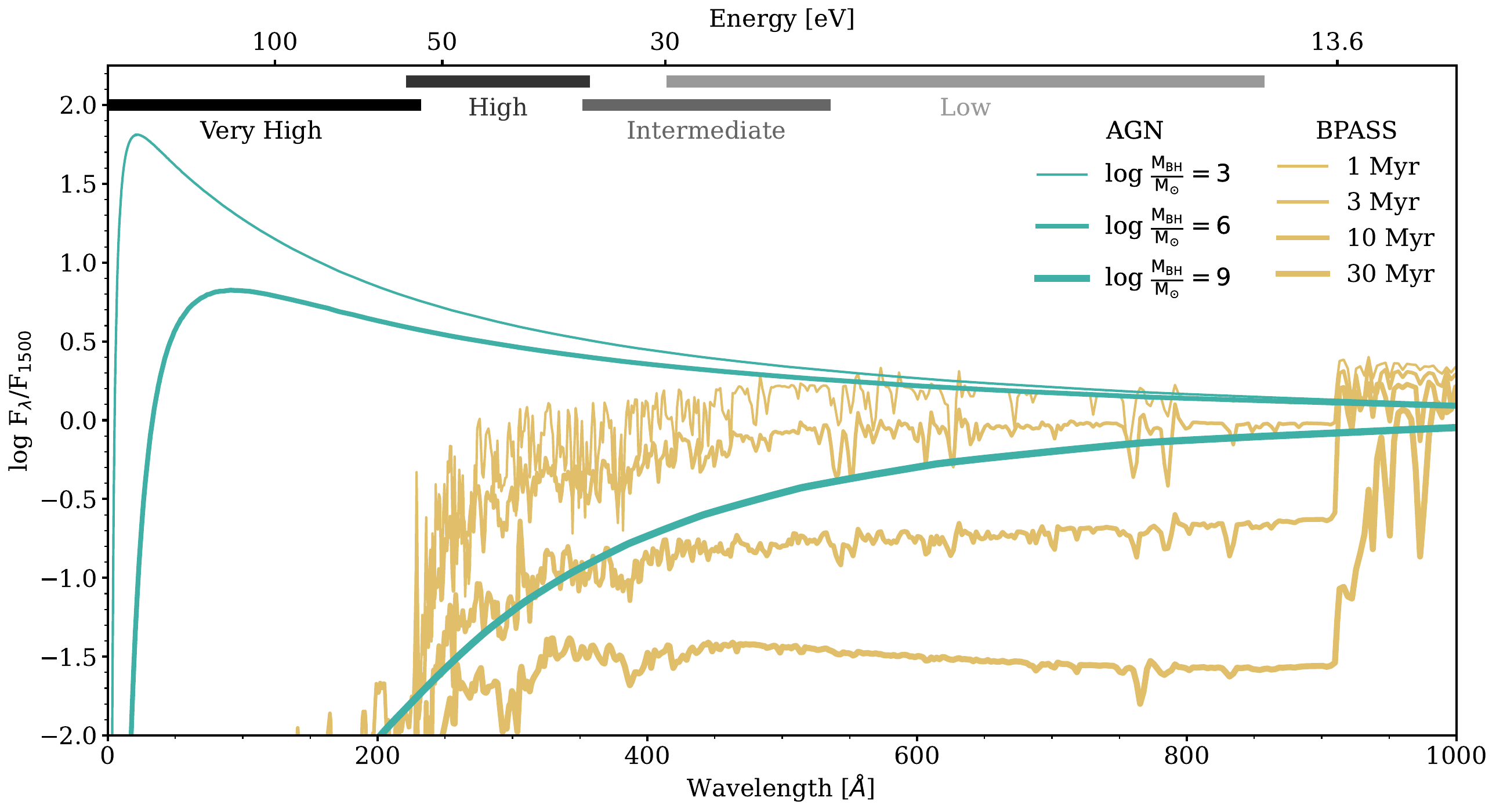}
\caption{Example ionizing spectra of four stellar population (gold) and three accreting black hole (blue) models used in this work. The annotations mark the four ``zones'' of ionization from \cite{Berg2021}. The stellar populations shown are constant initial mass function and metallicity with varying age, and the black hole accretion disk models are shown with constant $L_{bol}/L_{Edd}$ and spins and varying black hole masses. The full details of the models are discussed in Section \ref{sec:models}.
\label{fig:sed}}
\end{figure*}

\begin{figure*}[t]
\epsscale{1.1}
\plotone{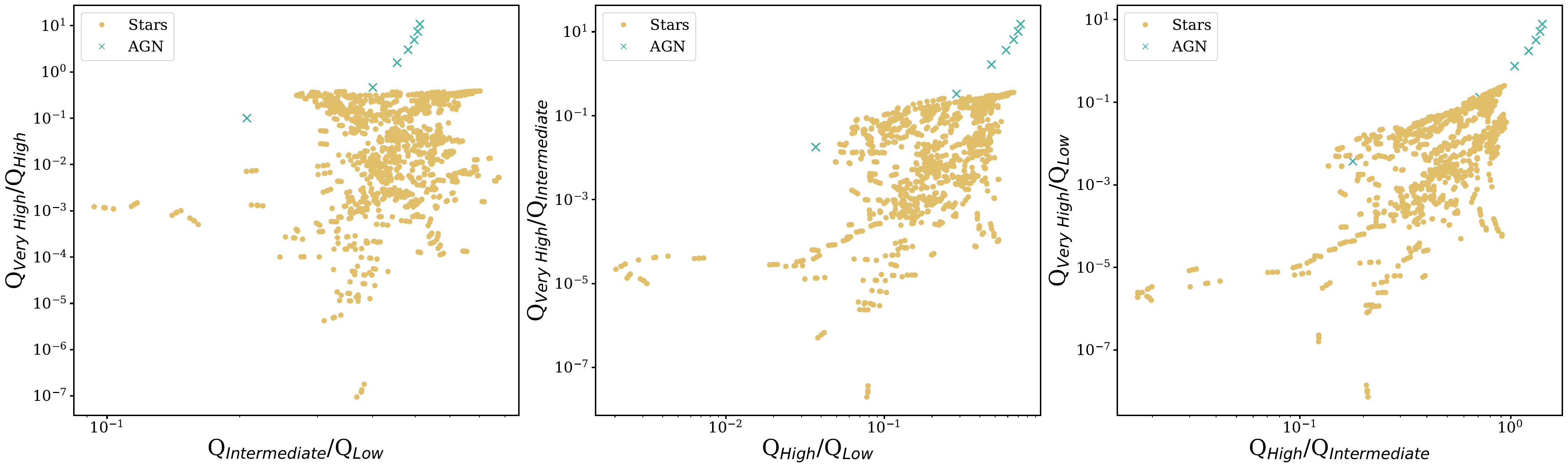}
\caption{The ratios of ionizing photon production $Q$ in each of the \cite{Berg2021} zones of ionization shown in Figure \ref{fig:sed} for all of the BPASS stellar populations and black hole accretion disk ionizing spectra used in this work. We show here the stellar models of log age/yr $<8$, excluding the ages for which there is no significant production of photons at the energies traced by the \nii-BPT, VO87, and OHNO diagrams.
\label{fig:sed_q}}
\end{figure*}

\section{Results}\label{sec:results}

\begin{figure*}[t]
\epsscale{1.15}
\plotone{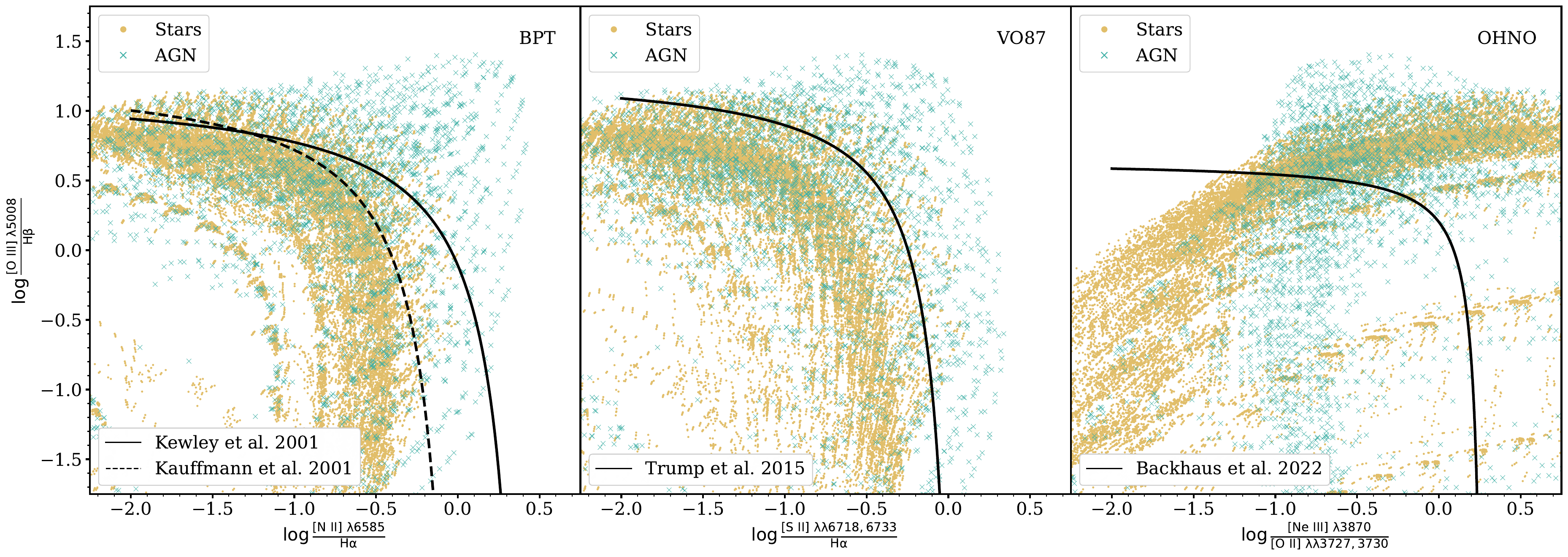}
\caption{The \nii-BPT (left), VO87 (center) and OHNO (right) diagrams with our suite of Cloudy models for BPASS stellar populations (gold circles) and black hole accretion disks (blue crosses). The lines show the demarcations for star formation and AGN (below and above the lines, respectively) for the \nii-BPT, \citep{Kauffmann2003,Kewley2001}, VO87 \citep{Trump2015}, and OHNO \citep{Backhaus2022}, respectively. The full model parameters are described in Section \ref{sec:models}. 
\label{fig:diagnostics_threepanel}}
\end{figure*}

\subsection{Analysis of Ionizing Spectra}
We first examine the stellar population and black hole accretion disk ionizing spectra before being processed through Cloudy. Figure \ref{fig:sed} shows several example ionizing spectra, including three black hole accretion disk models of varying black hole masses ($\log\mbh/\msun = 3,6,9$) and four different BPASS stellar populations of varying ages (age$=1,3,10,30$ Myr). We show the four zones of ionization from \cite{Berg2021}, which are defined as the ionization energies required to produce the following species:

\begin{enumerate}
    \item Low Ionization: N$^+$ (14.53-29.60 eV)
    \item Intermediate Ionization: S$^{2+}$ (23.33-34.79 eV)
    \item High Ionization: O$^{2+}$ (35.11-54.93 eV)
    \item Very High Ionization: He$^{2+}$ (>54.42 eV)
\end{enumerate}

Integrating the ionization spectra over the ionization zones gives the ionizing photon budget in each zone, which we denote \qtotal\ (integral under the entire H-ionizing continuum), \qlow, \qint, \qhigh, and \qveryhigh, respectively.

Figure \ref{fig:sed_q} shows the ratios of the ionizing photon budgets in each zone for the model stellar population and black hole accretion disk ionizing spectra. We choose to shown only the stellar models with ages log age/yr $\leq8$, as the older stellar populations in BPASS do not produce appreciable amounts of ionizing photons at energies traced by the \nii-BPT, VO87, and OHNO diagrams \citep{Stanway2018}.

\subsection{Inference of Model Properties from Emission Line Ratios}

Figure \ref{fig:diagnostics_threepanel} shows the three rest-optical emission line ratios which are the focus of this work: the \nii-BPT, VO87, and OHNO diagrams. On these diagrams we show a truncated set of our BPASS stellar populations and black hole accretion disk models, on which we perform the following analyses. 

Here we perform a simplistic Bayesian inference of the ionization parameter, gas-phase metallicity, and the ionizing photon budgets in each zone of the ionizing spectra. We calculated the log of the likelihood as a $\chi^2$, i.e.
\begin{align}
    \log P(\mathrm{data}|\theta) &\propto -\sum_i \frac{(\mathrm{data} - \mathrm{model_i})^2}{\mathrm{uncertainty}^2}
\end{align}

\begin{figure*}[t]
\epsscale{1.1}
\plotone{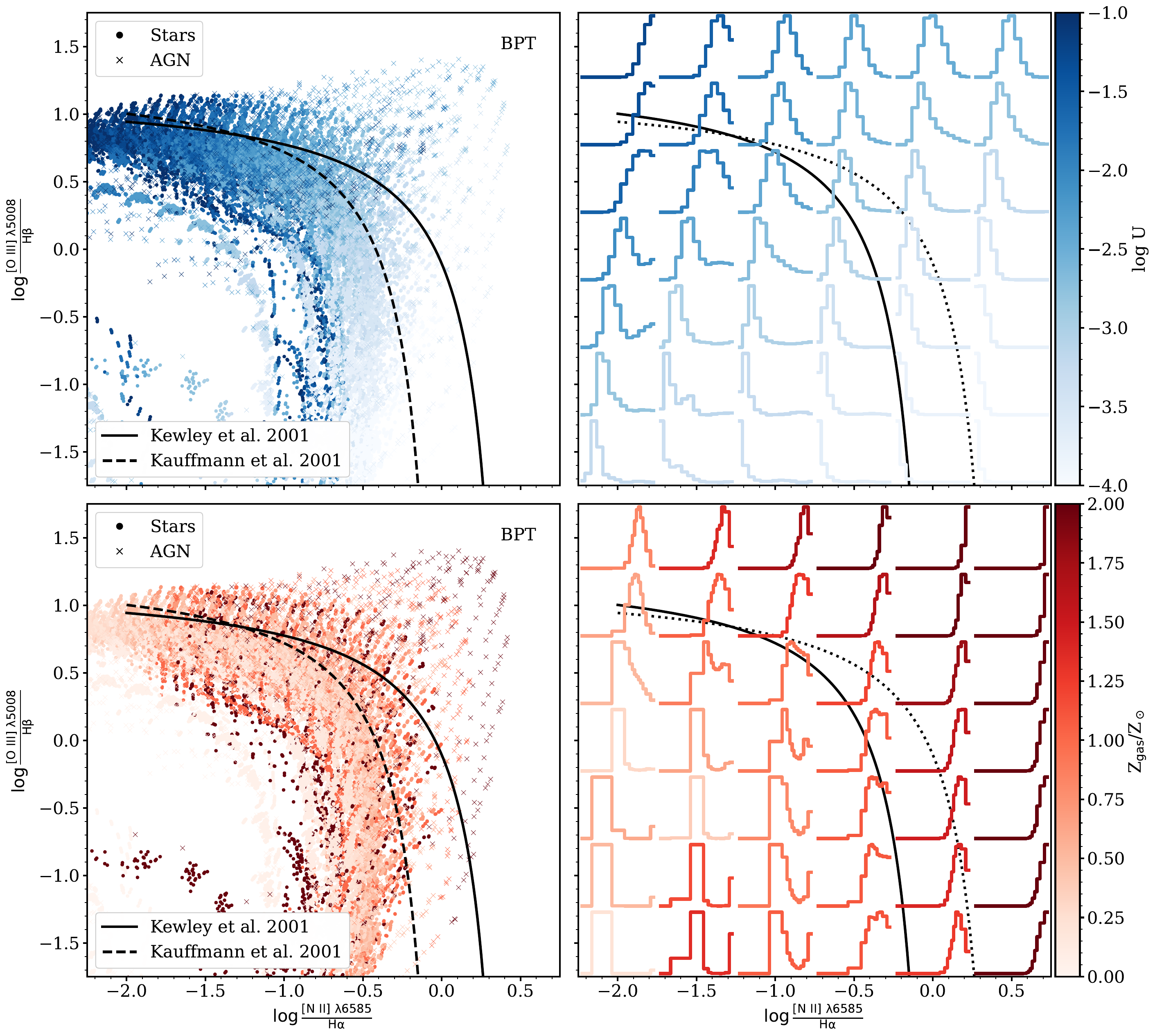}
\caption{The \nii-BPT diagram shown with our suite of photoionization models colored by ionization parameter (top left) and gas-phase metallicity (bottom left). The right panels show the posteriors $P(\theta|\mathrm{data})$ for ionization parameter (top right) and gas-phase metallicity (bottom right) for a mock observation at the respective point in the \nii-BPT plane, where the color shows the mean of the posterior. 
\label{fig:BPT_inference}}
\end{figure*}

\begin{figure*}[t]
\epsscale{1.1}
\plotone{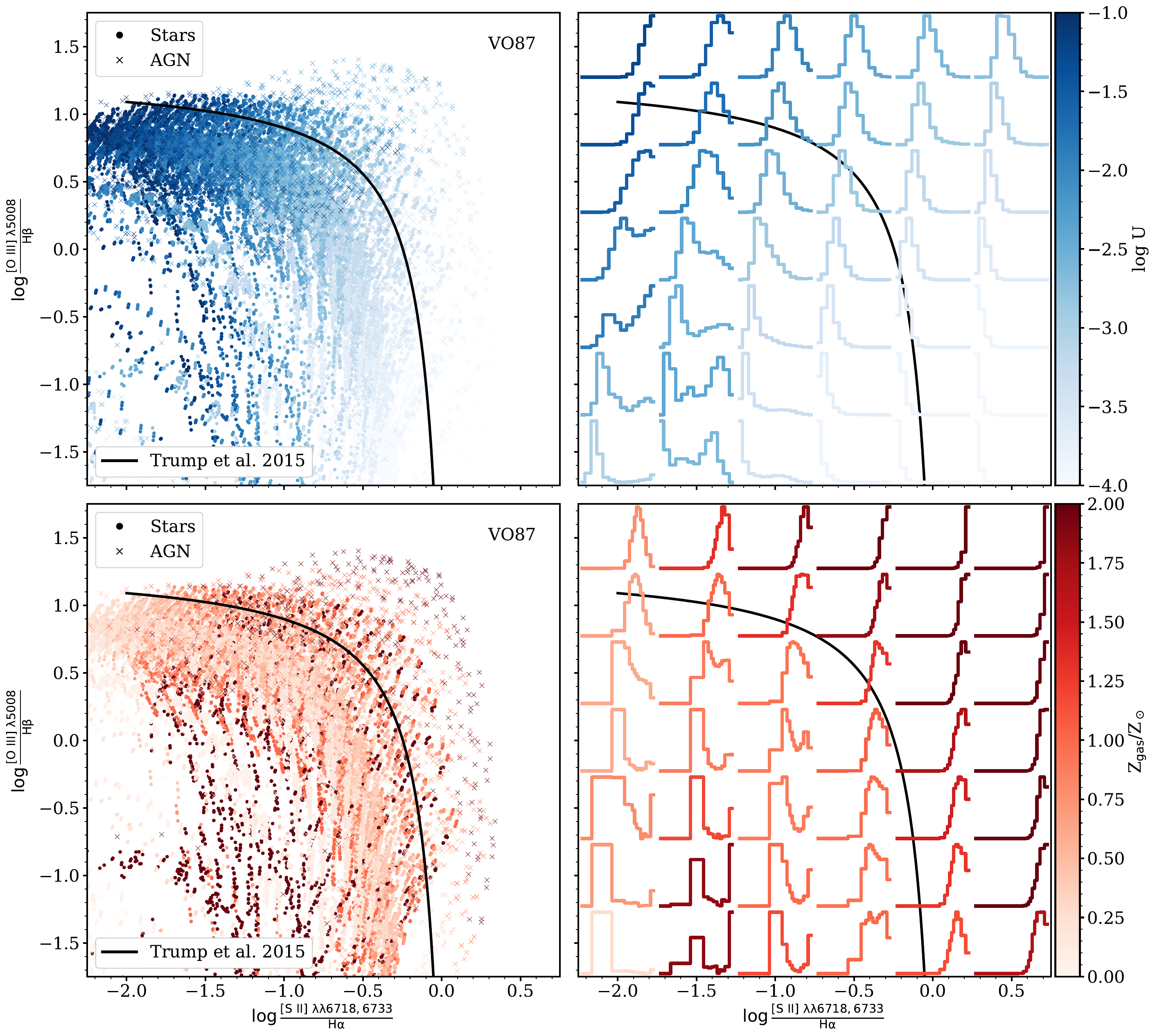}
\caption{The VO87 diagram shown with our suite of photoionization models colored by ionization parameter (top left) and gas-phase metallicity (bottom left). The right panels show the posteriors $P(\theta|\mathrm{data})$ for ionization parameter (top right) and gas-phase metallicity (bottom right) for a mock observation at the respective point in the VO87 plane, where the color shows the mean of the posterior. 
\label{fig:VO87_inference}}
\end{figure*}

\begin{figure*}[t]
\epsscale{1.1}
\plotone{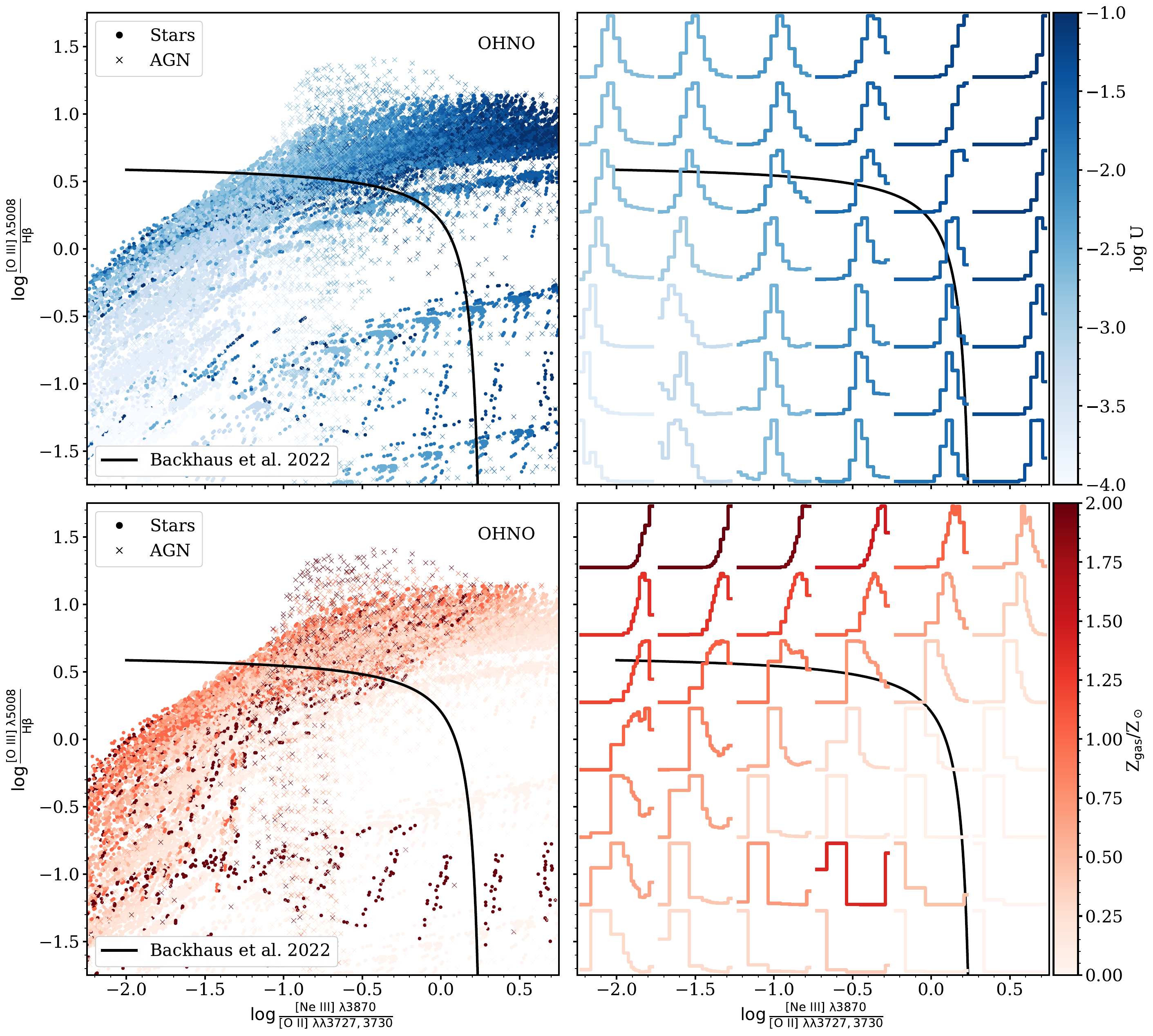}
\caption{The OHNO diagram shown with our suite of photoionization models colored by ionization parameter (top left) and gas-phase metallicity (bottom left). The right panels show the posteriors $P(\theta|\mathrm{data})$ for ionization parameter (top right) and gas-phase metallicity (bottom right) for a mock observation at the respective point in the OHNO plane, where the color shows the mean of the posterior. 
\label{fig:OHNO_inference}}
\end{figure*}

We perform this likelihood calculation over an evenly spaced grid (0.5 dex step size in each line ratio) in the \nii-BPT, VO87, and OHNO planes. For eight model parameters (log U, \zgas, and the ratios of the ionizing photon budgets parameterized as $Q_i/\qtotal$, where $i$ indexes over the four ionization zones, we invoke uniform priors to calculate the posteriors shown in Figures \ref{fig:BPT_inference}, \ref{fig:VO87_inference}, and \ref{fig:OHNO_inference}. We adopt the uncertainties to be half of the respective grid spacing (i.e., 0.25 dex in the relevant line ratio).

Figures \ref{fig:BPT_inference}, \ref{fig:VO87_inference}, and \ref{fig:OHNO_inference} shows the \nii-BPT, VO87, and OHNO diagrams, respectively, with photoionization models color-coded by ionization parameter and gas-phase metallicity. We also show the posteriors from the inference of the ionization parameter and gas-phase metallicity, with the likelihoods calculated in a discrete grid of 0.5 dex spacing in the respective line ratio plane. 

\begin{figure*}[t]
\epsscale{1.1}
\plotone{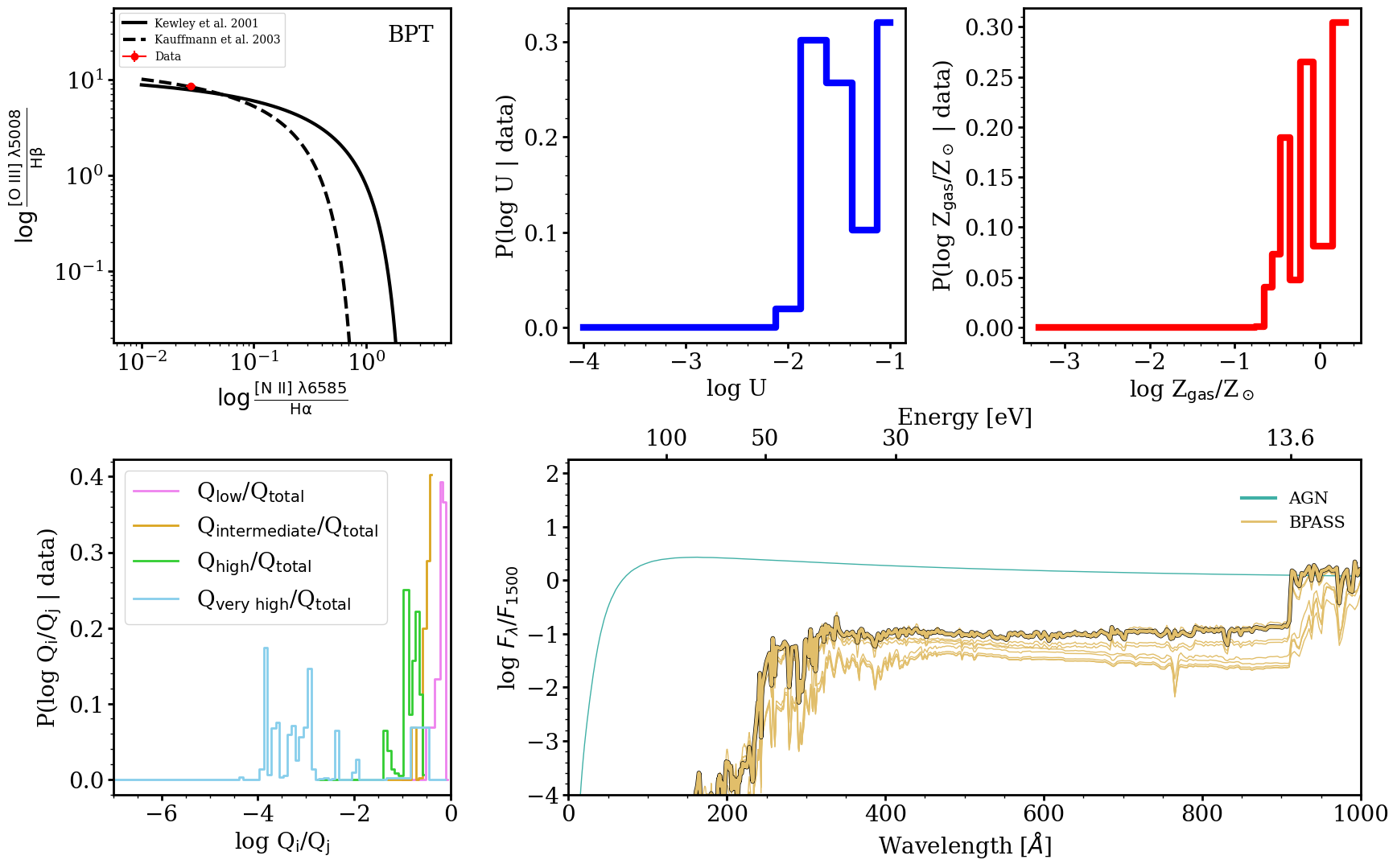}
\caption{The inferred properties from our models for an observation in the \nii-BPT plane, using the broad-line AGN GS\_3073 from \cite{Ubler2023} as an example. The top left panel shows the \cite{Ubler2023} object with 1$\sigma$ uncertainties in the \nii-BPT plane. The top center, top right, and bottom left panels show the inferred ionization parameter, gas phase metallicity, and ratios of the ionizing photon budgets in each ionization zone, respectively. The bottom right panel shows the ionizing continua of the ten most likely models at that point in the \nii-BPT plane, which include both accreting black holes (blue) and stellar populations (gold). We show the ionizing continuum of the model with the highest likelihood in bold. \label{fig:data_inference_bpt}}
\end{figure*}

Figure \ref{fig:data_inference_bpt} shows the inferred properties of an observation in the \nii-BPT plane. We choose the source GS\_3073 from \cite{Ubler2023} as a test observation, as it sits in the region of the \nii-BPT diagram that has significant overlap between black hole accretion disk and stellar population models at low to moderate gas-phase metallicities and high ionization parameters. GS\_3073 was observed in the NIRSpec IFS GTO program ``Galaxy Assembly with NIRSpec IFS'' (Program ID: 1216, PI: Nora L\"utzgendorf). GS\_3073 was observed with with the $R\sim1900-3600$ G395H/F290LP grating/filter pair on JWST/NIRSpec for a 5 hour integration time. GS\_3073 also has well-detected broad Balmer lines, which serve as an orthogonal measure in support of the presence of an accreting massive black hole. We show the posteriors for the ionization parameter, gas-phase metallicity, and ionizing photon budget ratios. We also show the ionizing continua for the 10 most likely models, which include both young stellar populations and accreting black holes.

\begin{figure*}[t]
\epsscale{1.1}
\plotone{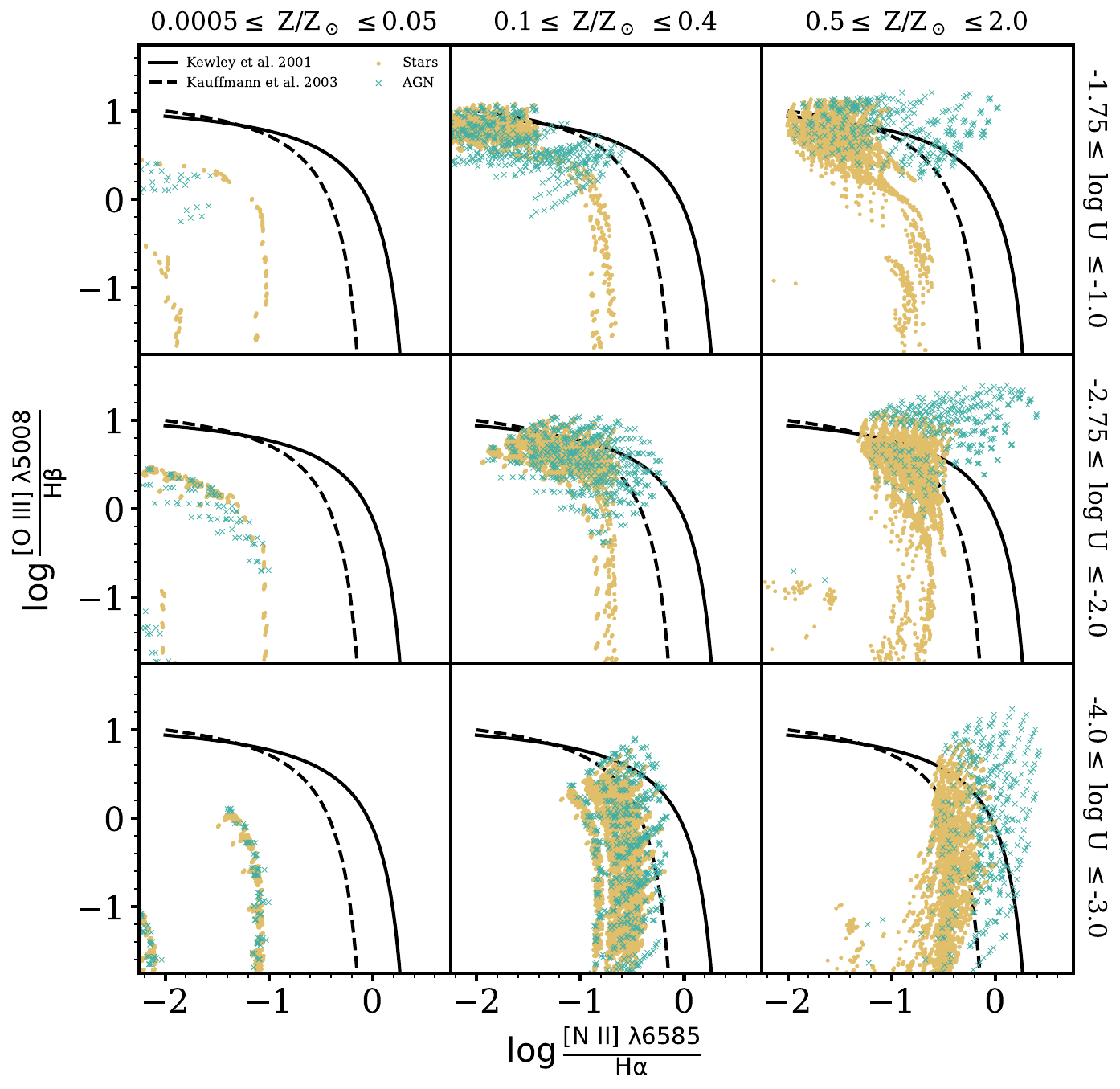}
\caption{The \nii-BPT diagram with our suite of Cloudy models for BPASS stellar populations (gold circles) and black hole accretion disks (blue crosses) in slices of ionization parameter (increasing up) and gas-phase metallicity (increasing to the right). The lines show the demarcations for star formation and AGN (below and above the lines, respectively) for the \nii-BPT diagnostics from \cite{Kauffmann2003} (dashed) and \cite{Kewley2001} (solid). We also include an animated version of the \nii-BPT, VO87, and OHNO diagrams in steps of ionization parameter and metallicity\footnote{The animated versions of Figures \ref{fig:bpt_grid}, \ref{fig:vo87_grid}, and \ref{fig:ohno_grid} are  available here:  \href{https://github.com/njcleri/AR_05558_modeling/tree/main/figures/animations}{https://github.com/njcleri/AR\_05558\_modeling/tree/main/figures/animations}}.
\label{fig:bpt_grid}}
\end{figure*}

\begin{figure*}[t]
\epsscale{1.1}
\plotone{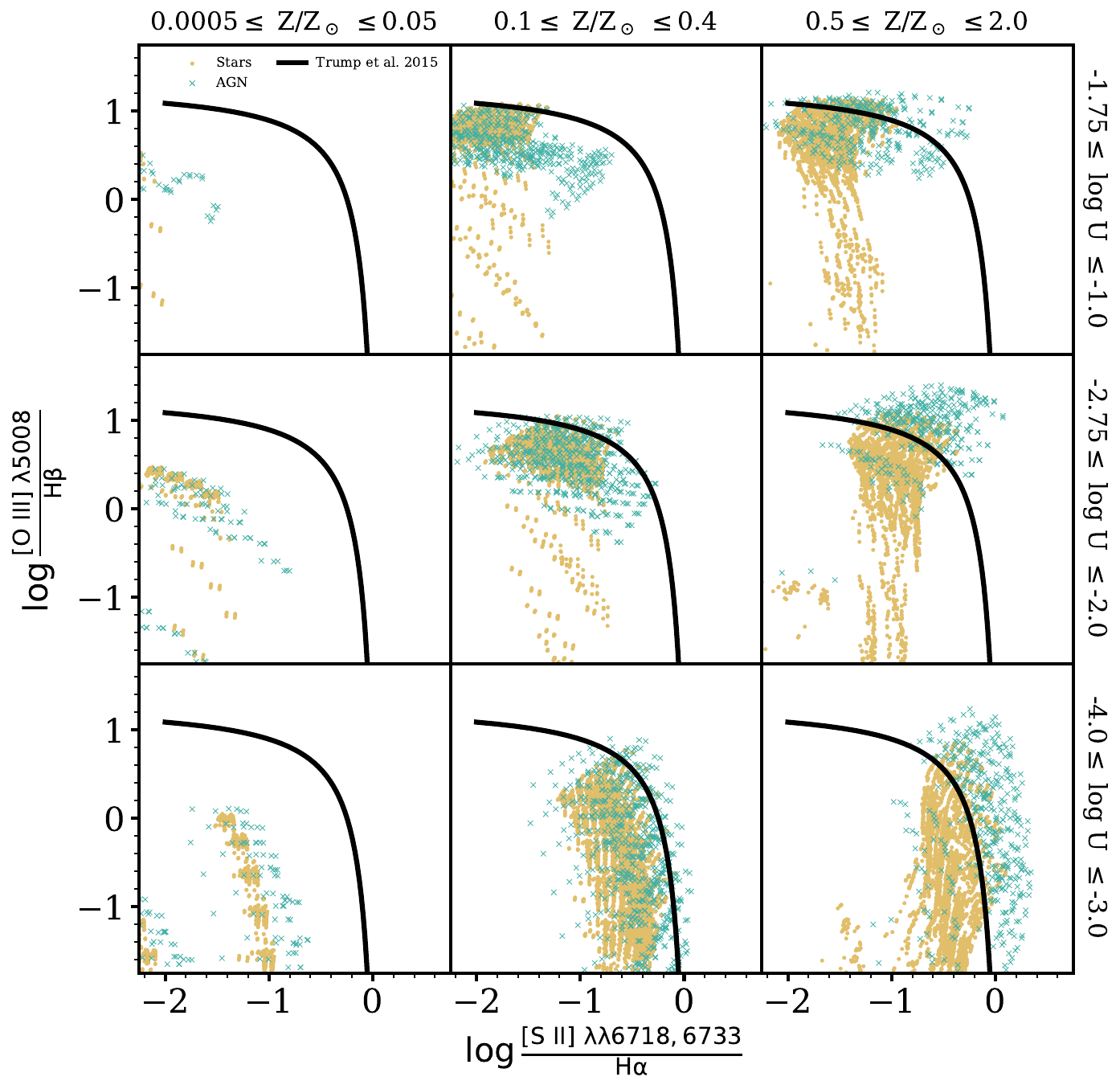}
\caption{The VO87 diagram with our suite of Cloudy models for BPASS stellar populations (gold circles) and black hole accretion disks (blue crosses) in slices of ionization parameter (increasing up) and gas-phase metallicity (increasing to the right). The lines show the demarcations for star formation and AGN (below and above the lines, respectively) for the VO87 diagnostic from \citep{Trump2015}.
\label{fig:vo87_grid}}
\end{figure*}

\begin{figure*}[t]
\epsscale{1.1}
\plotone{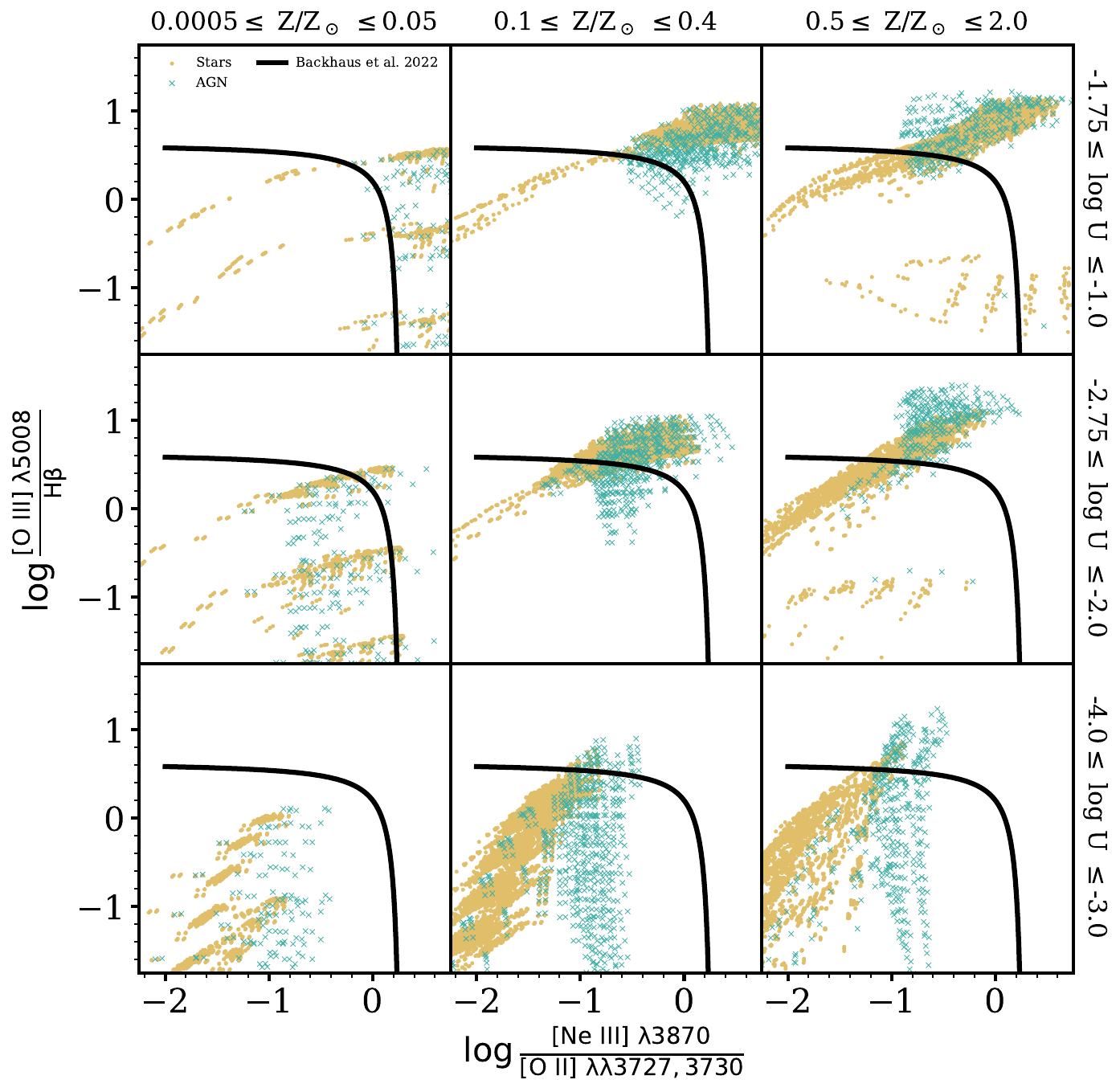}
\caption{The OHNO diagram with our suite of Cloudy models for BPASS stellar populations (gold circles) and black hole accretion disks (blue crosses) in slices of ionization parameter (increasing up) and gas-phase metallicity (increasing to the right). The lines show the demarcations for star formation and AGN (below and above the lines, respectively) for the OHNO diagnostic from \cite{Backhaus2022}.
\label{fig:ohno_grid}}
\end{figure*}

\section{Discussion}\label{sec:discussion}
\subsection{Constraints on Nebular Conditions}
Figure \ref{fig:diagnostics_threepanel} shows the \nii-BPT, VO87, and OHNO diagrams with a truncated set of Cloudy models discussed in Section \ref{sec:models}. We show the extent of the stellar and accretion disk models in this parameter space, along with the impact of the dimensionless ionization parameter and the gas phase metallicity. We show that the \nii-BPT and VO87 diagrams do have a separation between the stellar and accretion disk models when marginalized over all model parameters, though not necessarily in line with the \cite{Kewley2001}, \cite{Kauffmann2003}, and \cite{Trump2015} diagnostics. The disagreement here is likely in the construction of the model grids in this work; we probe a very broad parameter space of stellar population models, as well as a large spread of ionization parameters and densities, designed to include the maximal extents of ``normal'' stellar populations potentially beyond those of the original \cite{Kewley2001,Kewley2006} diagnostics (see Section \ref{sec:models}). The OHNO diagram shows that the stellar models follows tracks of increasing ionization parameter across the \cite{Backhaus2022} diagnostic, with a similar extent as the black hole accretion disk models. 

Figure \ref{fig:BPT_inference} shows the inference of the nebular parameters in the \nii-BPT diagram. This analysis reaffirms the intended physical interpretations of these ratios: ionization parameter increases with increasing \oiii/\hb\ and the gas-phase metallicity increases with increasing \nii/\ha. Figure \ref{fig:VO87_inference} shows similar behavior in the VO87 diagram, where ionization parameter increases with increasing \oiii/\hb\ and the gas-phase metallicity increases with increasing \sii/\ha. 

Figure \ref{fig:OHNO_inference} shows the inference of the ionization parameter and nebular metallicity in the OHNO diagram. We see that the position on the OHNO diagram is most strongly driven by the ionization parameter. This is primarily due to the ionization energies needed to produce the respective emission lines: 35.12 eV and 13.62 eV for \oiii\ and \hb, and 40.96 eV and 13.62 eV for \neiii\ and \oii. These axes are near-completely degenerate in ionization energies probed; both \oiii/\hb\ and \neiii/\oii\ trace the ``high'' to ``low'' ionization ratio of the ionizing spectra, as shown in Figure \ref{fig:sed}. 

We show these behaviors explicitly in Figures \ref{fig:bpt_grid}, \ref{fig:vo87_grid}, and \ref{fig:ohno_grid} (and in the linked animations), which show the \nii-BPT, VO87, and OHNO diagrams in a grid of ionization parameter and metallicity. 

The \nii-BPT and VO87 diagrams behave generally as predicted \citep[e.g.,][]{Baldwin1981,Veilleux1987,Kewley2001,Kewley2006,Kewley2013,Kewley2019b,Kauffmann2003}: higher ionization parameter is consistent with higher \oiii/\hb\ and lower \nii/\ha\ or \sii/\ha, and metallicity is less constrained but higher metallicity is generally more consistent with higher \nii/\ha\ or \sii/\ha. Unfortunately, Figures \ref{fig:bpt_grid} and \ref{fig:vo87_grid} show that there is still significant overlap and contamination of both the AGN and star forming regions of the \nii-BPT and VO87 diagrams, particularly at lower metallicity and higher ionization parameter. 

The OHNO diagram takes on a different behavior than the \nii-BPT and VO87 diagrams. The bulk of the models in the OHNO parameter space (see Figures \ref{fig:OHNO_inference} and \ref{fig:ohno_grid}) move into the AGN region at high ionization parameter (log U $\gtrsim-2$), regardless of the stellar or black hole driven ionizing spectrum. This ionization parameter-driven contamination makes the OHNO diagram worse at separating stellar populations from accreting black holes as redshift increases. This is consistent with the interpretation of the OHNO diagram being a better diagnostic at the low redshifts for which it was first constructed ($z\sim1$), where the star-forming regions tend to have lower ionization parameters than at the much higher redshifts probed by JWST \citep[e.g.,][]{Backhaus2022,Backhaus2024,Trump2023,Cleri2023b,Larson2023,Kocevski2023}.

An additional complication to this is the potential disjointed evolution and distinct nebular physics of stellar populations and accreting black holes across cosmic time, which would lead to greater scatter than what is presented in Figures \ref{fig:bpt_grid}, \ref{fig:vo87_grid}, and \ref{fig:ohno_grid}. This motivates a systematic analysis of the optical strong line ratios of high redshift observations (Cleri et al. in preparation). We discuss the complications of redshift evolution in Section \ref{subsec:high_redshift}.

\subsection{Constraints on the Ionizing Continuum}
The critical function of line ratio diagnostics, and AGN selection as a whole, is to discern the source(s) of the ionizing continuum from the respective input data. 

Figure \ref{fig:sed_q} shows the ratios of the ionizing photon budgets in each of the \cite{Berg2021} ionization zones. Considering the context of the line ratios used in the \nii-BPT, VO87, and OHNO diagrams, the ionization zones traced by the respective line ratios are high/low (\oiii/\hb, \neiii/\oii) and low/low (\nii/\ha, \sii/\ha). Strictly from the ionizing continuum, there is substantial overlap between the stellar population and black hole accretion models in the \qhigh/\qlow\ regime, indicating that the shapes of the ionizing continua are similar at these energies. This is shown in that the stellar population models overlap significantly with the black hole accretion disk models in the \qhigh/\qlow, particularly at black hole masses in the SMBH regime (log \mbh/\msun $\geq$ 6), which are the most commonly observed black holes. These models offer a solution not explored in the \nii-BPT, VO87, and OHNO diagrams in the form of very-high-ionization emission. Emission lines that require >54.42 eV (e.g., \heii, \nev) break most of the degeneracy immediately. This is shown clearly in Figure \ref{fig:data_inference_bpt}, where the most likely ionizing continua may be similar in the $<54$ eV regime but significantly different at higher energies. 

Ionization-based diagnostics (e.g., \oiii/\hb) are more directly tied to the shape of the ionizing continuum than ionization-insensitive nebular tracers (e.g., \nii/\ha). By extension, adding another anchor at higher ionization (e.g., \heii\ or \nev) gives a direct constraint on the shape of the ionizing spectrum than the addition of ionization-insensitive constraints. Thus, the addition of information about the production of >54 eV photons is the strongest optical constraint on the ionizing source. This is consistent with the conclusions of other works which study >54 eV emission lines as a strong tracer of black hole accretion \citep[e.g.,][]{DeBreuck2000,Abel2008, Petric2011,Cleri2023a,Cleri2023b,Chisholm2024}.

Figure \ref{fig:data_inference_bpt} shows that it is difficult to uniquely constrain the source of the ionizing continuum for real observations from the \nii-BPT diagram alone. We show that even with the high quality deep spectroscopy from GA-NIFS, the region of the \nii-BPT diagram where GS\_3073 falls hosts too significant of an overlap between the BPASS stellar populations and accreting black hole models to provide a unique solution for the source of the ionizing continuum. This is true even with the density of stellar models being much higher than that of the accreting black hole models. 

\subsection{Caveats and Limitations}\label{subsec:caveats}
The results and conclusions of this work are subject to several caveats and limitations, which we discuss here. 

An inherent issue in photoionization modeling is the sampling of parameter space. It is not expected that the full complexity of nebular physics in a galactic environment can be described in such a small number of model parameters, or that the assumptions made about aspects of the ionized medium (e.g., geometry, solar abundance scaling, etc.) are generalizable to all nebulae \citep[see, e.g.,][for recent efforts toward more generalized photoionization simulations]{Katz2023c}. 

The stellar population model grids computed for this work were designed to cover a very broad range of stellar population properties (e.g., IMF, age, stellar metallicity, binaries) and nebular parameters (e.g., ionization parameter, gas phase metallicity) with the intent of finding the maximal extent of ``normal'' stellar populations (e.g., not Population III stars, high-mass X-ray binaries, etc.; see Section \ref{subsec:other_sources} for further discussion) in line ratio diagnostic space.

\subsubsection{Models of Black Hole Accretion}
Perhaps the most obvious limitation of this work is the simplicity of the black hole accretion disk models. Black hole accretion physics is multivariate, with many physical parameters that can be changed which each affect the ionizing continuum. In this work we vary the black hole mass, where, all else constant, increases in black hole mass lead to cooler/softer ionizing continua. Additionally, the accretion rate and black hole spin can each change the shape and normalization of the ionizing continuum. Combining these three parameters can even give extreme cases where the accretion disk gives no ionizing photons \citep[in the case of cold lineless quasars, e.g.,][]{Laor2011,Hryniewicz2010}.

Many often-used accretion disk SED models are some permutation of a broken power law, such as that from \cite{Mathews1987}\footnote{The AGN SED default in Cloudy (accessed via the \texttt{table agn} command) differs from the \cite{Mathews1987} template only in the 10 \micron\ break, which does not affect the analyses of this work (see the version C17.01 \citealt{Ferland2017} documentation for more information).}, similar to a typical radio quiet AGN. In the optical, this continuum corresponds to a simple power law with slope $\alpha=-1$.  

Other often-used models include those of well-studied local active galaxies, including the nearby Seyfert 1 galaxy NGC 5548 \citep{Mehdipour2015}. NGC 5548 has a multiwavelength SED from near-infrared to the hard (200 keV) X-ray. The UV to near-IR continuum of NGC 5548 is consistent with a single Comptonized disk component, with no evidence of an additional purely thermal disk component or additional component of reprocessing from the disk. NGC 5548 also exhibits a ``soft X-ray excess'' often seen in local AGN \citep[e.g.,][]{Mehdipour2015,Done2012}. NGC 5548 has a central engine powered by a $6.5\times10^7$\msun\ supermassive black hole \citep{Bentz2007}, and the SED has a shape consistent with a black hole of this mass in the models described in Section \ref{sec:models} of this work \citep[see also][]{Cleri2023b}. 

Several works suggest other empirical SED shapes for low-redshift Seyferts \citep{Binette1989,Clavel1990} or high-redshift quasars \citep{Zheng1997,Korista1997,VandenBerk2001,Richards2006,Fan2006}, yet these models are all broadly similar in shape to the \citealt{Done2012} SEDs. There also exist many AGN which are not well described by the \citealt{Done2012} SEDs or similar models, including low-luminosity AGN \citep[e.g.,][]{Ho2008}, and the recently discovered ``little red dots'' at higher redshifts from JWST spectra \citep[e.g.,][]{Matthee2024,Hviding2025,Kocevski2023,Kocevski2025,Taylor2024,Taylor2025,Wang2024,Wang2025}.

\subsubsection{Other Sources of Ionizing Photons}\label{subsec:other_sources}
The model grid used in this work is limited to BPASS stellar populations and black hole accretion disk models. There exist many other potential sources of ionizing radiation which only serve to add to the confusion in these optical strong line ratio diagnostics.

Shocks represent the most confounding source of emission line production not explored in this work \citep[e.g.,][]{Ferland1983,McKee1980,Draine1993,Dopita1995}. Shocks have many origins, including mergers \citep[e.g., ][]{Medling2015}, stellar winds associated with Wolf-Rayet stars \citep[e.g., ][]{Simpson2007}. Stellar winds driven by starbursts are often observed in higher redshift galaxies with higher star formation rates than galaxies in the low redshift Universe \citep[e.g., ][]{Weiner2009,Steidel2010,Izotov2012,Izotov2021,Chisholm2018,Rigby2018,Davies2019,Davies2024}. Shocks with velocities of several hundred kilometers per second can be produced by a combination of starburst driven winds and supernovae or from outflows originating from an accreting black hole \citep[e.g.,][]{Evans1999,Veilleux2005,Veilleux2023,Vayner2023}. Shocks have been modeled extensively \citep[e.g.,][]{Thuan2005,Dopita1995,Izotov2012,Izotov2021}, and have been shown to cover a large parameter space of optical line ratio diagnostics \citep[e.g., the \nii-BPT and VO87 diagrams,][]{Dopita1995}.

Accreting stellar mass compact objects, e.g., X-ray binaries, are another potential source of high-energy ionizing photons \citep[e.g.,][]{Garofali2024}. The impact of high-mass X-ray binaries (HMXBs) on the ionizing photon budget of a galaxy and the observed spectrum has been disputed \citep[e.g.,][]{Schaerer2019,Senchyna2017}, though HMXBs may play an important role in the production of higher energy emission features, particularly \heii\ \citep[e.g.,][]{Shirazi2012,Jaskot2013}. 

Throughout this work, we have demonstrated that there exists significant confusion in the rest-frame optical line ratio diagnostics when considering only BPASS stellar populations and black hole accretion disk models. The consideration of other sources of ionizing photons serves to add to the confusion in this parameter space; this indicates that great care should be taken when using these diagrams, beyond what our analysis alone shows. 

\subsection{Line Ratio Diagnostics at High Redshifts}\label{subsec:high_redshift}

A complete understanding of line ratio diagnostics requires understanding the coevolution of stars and black holes along with the gas-phase conditions of galaxies across cosmic time. Pre-JWST observations and simulations indicate harder ionizing spectra, increases in ionization parameter, electron density, star formation rate, and decreases in metallicity in predominantly star-forming galaxies out to moderate redshifts \citep[e.g.,][]{Hainline2009,Bian2010,Bian2016,Siana2010,Kewley2013,Madau2014,Shapley2015,Shapley2019,Shivaei2015,Shivaei2018,Sanders2015,Sanders2016,Sanders2018,Sanders2020,Steidel2014,Steidel2016,Strom2017,Strom2018,Strom2022,Suzuki2017,Kaasinen2017,Kashino2017,Gburek2019,Maiolino2019,Simons2021,Papovich2022,Backhaus2022}. JWST-era studies have indicated that many of these trends in metallicity, electron temperature, electron density, star formation rates, and ionizing photon production continue out to much higher redshifts \citep[e.g.,][]{Isobe2023,Sanders2023,Sanders2024,Abdurrouf2024,Backhaus2024,Cameron2023a,Christensen2023,Curti2023,Fujimoto2023a,Heintz2023,Hsiao2024a,Hsiao2024b,Nakajima2025}; therefore, we find it reasonable to assume that the high-redshift Universe is broadly consistent with lower metallicities and higher ionization parameters.

The evolution of accreting black hole demographics across cosmic time presents another complication to this field. Unfortunately, observations of high-redshift accreting black holes are not necessarily constraining to the physical parameters which drive the ionizing continuum (e.g., black hole mass, spin, accretion rate). Black hole mass estimates (and by extension, Eddington ratios) are highly uncertain at high redshifts, primarily limited to kinematics of broad permitted lines from a single epoch \citep[e.g.,][]{Greene2005,Harikane2023,Kokorev2023,Larson2023,Maiolino2024b,Taylor2024,Taylor2025,Kocevski2023,Kocevski2025,Sun2025,Matthee2024,Hviding2025}. Unfortunately, less systematics-dominated methods such as reverberation mapping \citep[e.g.,][]{Blandford1982,Peterson1993,Peterson2004,Kaspi2000} and gas or stellar dynamics \citep[e.g.,][]{Ferrarese2000,Gebhardt2000,Kormendy2013} are predominantly not accessible beyond the local Universe with current instrumentation \citep[e.g.,][]{Abuter2024,Golubchik2024,Pacucci2024,Cohn2025,Newman2025}. 

The \nii-BPT and VO87 diagrams have been shown to be highly effective at separating star-forming galaxies from accreting supermassive black hole hosts in local studies \citep[e.g.,][]{Baldwin1981,Veilleux1987,Kauffmann2003,Kewley2006,Kewley2019b,Trump2015}. However, the local calibrations of these diagnostics may not hold at $z\lesssim2$ as indicated by pre-JWST studies \citep[e.g.,][]{Baldwin1981,Veilleux1987,Kewley2006,Juneau2011,Juneau2014,Coil2015}, and have been further called into question at $z>2$ as shown by recent studies with JWST spectroscopy \citep[e.g.,][]{Sanders2023,Ubler2023,Cameron2023a}. 

The OHNO diagram has also been used in several works in the early JWST era with observations of very high redshift galaxies \citep[up to $z\sim9$, e.g.,][]{Trump2023,Larson2023,Backhaus2024,Cleri2023b,Kocevski2023,Kumari2024,Hu2024,Gupta2024,Killi2024,Calabro2024,Wu2025,ArevaloGonzalez2025,Rinaldi2025,Scholtz2025,Treiber2025}. Several of these works have noted that nearly all high-redshift observations (particularly at $z\gtrsim5$) land in the AGN region of the $z\sim1$ \cite{Backhaus2022} diagnostic, calling into question the utility of OHNO at early epochs. 

Figures \ref{fig:bpt_grid}, \ref{fig:vo87_grid}, and \ref{fig:ohno_grid} show that the stellar and black hole models are poorly behaved in the \nii-BPT, VO87, and OHNO diagrams at moderate to high ionization parameters (log U $\geq$-2.75) and moderate to low metallicities (\zgas/\zsol\ $\leq$ 0.4). The OHNO diagram is particularly poorly behaved, with near-complete contamination of the AGN region at high ionization parameters. This indicates that more information beyond these two-dimensional line ratio diagnostics is needed to fully characterize a source at high redshifts where these gas conditions are increasingly common. 

\section{Summary and Conclusions}\label{sec:conclusions}
In this work, we employ a large photoionization model grid computed using Cloudy to study three optical emission line ratio diagnostics of ionizing sources: \oiii/\hb\ vs. \nii/\ha\ \citep[the ``\nii-BPT'' diagram][]{Baldwin1981}, \oiii/\hb\ vs. \sii/\ha\ \citep[the ``VO87'' diagram][]{Veilleux1987}, and \oiii/\hb\ vs. \neiii/\oii\ \citep[the ``OHNO'' diagram][]{Backhaus2022}. We analyze the ionizing spectra and the Cloudy models and perform parameter inference to predict the physical conditions of a source given its optical strong line ratios.   

The primary findings of this work are as follows: 
\begin{itemize}
    \item The optical emission line ratio diagnostics \nii-BPT, VO87, and OHNO are strong tracers of the ionization parameter and gas-phase metallicity. At moderate to high ionization parameters (log U $\geq$-2.75) and moderate to low metallicities (\zgas/\zsol\ $\leq$ 0.4), there is significant and sometimes near-complete overlap between the stellar population and black hole accretion disk models (see Figures \ref{fig:bpt_grid}, \ref{fig:vo87_grid}, and \ref{fig:ohno_grid}). 
    \item We show that the OHNO diagram in particular is most strongly driven by the ionization parameter log U, and is not necessarily sensitive to the physical source of the ionizing photons.  The contamination of the AGN region of the OHNO diagram is worse at higher ionization parameter, which makes OHNO a worse diagnostic at increasing redshifts. Thus we do not recommend the use of OHNO alone as a diagnostic of ionizing sources, particularly in the gas conditions typical of galaxies in the early Universe. 
    \item There is a significant overlap in the range of the black hole accretion disk and stellar ionizing continua when parameterized by the ratios of the ionizing photon budgets Q in each of the four ionization zones, with the exception of ratios with \qveryhigh\ (see Figure \ref{fig:sed_q}). This is evidence that spectral features which probe >54 eV photons (e.g., \heii, \nev, etc.) are strong indicators of an ionizing source harder than BPASS stellar populations and should be targets for future study.
    \item We show that it is difficult to uniquely constrain the dominant ionizing sources for real observations of high-redshift galaxies with optical strong line ratios alone (see Figure \ref{fig:data_inference_bpt}). 
\end{itemize}

Our results show that these three optical strong line ratio diagnostics, the \nii-BPT, VO87, and OHNO diagrams are highly convenient and useful due to the observability of the rest-frame optical strong lines in many redshift regimes. However, each of the three has underlying biases and confounding behavior which need to be carefully considered for their results to be meaningfully interpreted. 

There exist many outstanding questions about the determination of ionizing sources in galaxies at early times. Future works will test these optical diagnostics and many others with large statistical samples, now made accessible with JWST spectroscopy (e.g., Cleri et al. in preparation). The impacts of shocks, binary interactions, and other sources of ionization at high redshifts still remain unclear (see Section \ref{subsec:other_sources} for a discussion), and further serve to complicate the task of characterizing high-redshift observations. 

To fully understand the underlying physics which drives the emission from galaxies requires more information than a small number of strong emission line ratios can provide. While emission line ratio diagnostics from an integrated spectrum are very convenient tools to provide initial evidence of an ionization mechanism, we need to look beyond using a single piece of information to strictly dichotomize an astrophysical source. The results of this work motivate future observations of high redshift galaxies with deep spectroscopy from JWST and multiwavelength data from X-ray to radio in order to form a complete picture of the physical mechanisms driving the ionization conditions of galaxies in the early Universe. 


\software{Astropy \citep{Astropy2013}, NumPy \cite{Harris2020}, Matplotlib \citep{Hunter2007}}, Cloudy \citep{Ferland2017,Gunasekera2023}, pandas \citep{Reback2022}, \texttt{XSPEC} \citep{Arnaud1996}

\begin{acknowledgements}
NJC thanks the CEERS and RUBIES collaborations, Kartheik Iyer, and Joshua Speagle for insightful conversations throughout the course of this work. 

This work is primarily funded by the NASA grant JWST-AR-05558. NJC also acknowledges support from the Eberly Postdoctoral Fellowship in the Eberly College of Science at The Pennsylvania State University. Portions of this research were conducted with the advanced computing resources provided by Texas A\&M High Performance Research Computing (HPRC; http://hprc.tamu.edu). This work benefited from support from the George P. and Cynthia Woods Mitchell Institute for Fundamental Physics and Astronomy at Texas A\&M University.
\end{acknowledgements} 

\begin{contribution}
    NJC led the design and analysis of this work and the proposal from which this work is funded. GMO helped develop the photoionization models used throughout this work. BEB, JL, CP, and JRT offered high-level directional guidance throughout the work. All other authors provided useful commentary and feedback on the analysis, communication, and interpretation of the results in this work and in the writing of the proposal from which this science is funded, NASA grant JWST-AR-05558.
\end{contribution}

\clearpage
\bibliography{library}{}
\bibliographystyle{aasjournal}{}

\end{document}